\documentclass[twocolumn,showpacs,showkeys,preprintnumbers,amsmath,amssymb,prd]{revtex4}

\usepackage{graphicx}
\usepackage{dcolumn}
\usepackage{bm}
\begin{document}


\title{Quasigroups, Asymptotic Symmetries and Conservation Laws in General
Relativity }

\author{Alexander I. Nesterov}
\affiliation{Departament of Physics, C.U.C.E.I., Guadalajara University,
Guadalajara, Jalisco, Mexico}
\email{nesterov@cencar.udg.mx}

\date{\today}

\begin{abstract}
A new quasigroup approach to conservation laws in general relativity is
applied to study asymptotically flat at future null infinity spacetime. The
infinite-parametric Newman-Unti group of asymptotic symmetries  is
reduced to the Poincar\'e quasigroup and the Noether charge associated with
any element of the Poincar\'e quasialgebra is defined. The integral
conserved quantities of energy-momentum and angular momentum are linear on
generators of Poincar\'e quasigroup, free of the supertranslation
ambiguity, posess the flux and identically equal to zero in Minkowski
spacetime.
\end{abstract}

\maketitle

As well known the general covariance of the Einstein equations results in
differential conservation laws related to the field equations. From
Noether's theorem one shows that for an arbitrary diffeomorfism generated
by the vector field $\xi$ the invariance of the Lagrangian for the Einstein
equations leads to the conservation laws of the form
\begin{equation}
{\frak J}^\mu(\xi)_{,\mu}=0,
\end{equation}
where the vector density ${\frak J}^\mu$ is defined as
${\frak J}^\mu(\xi)={\frak h}^{\mu\nu}_{,\nu}$,
${\frak h}^{\mu\nu}=-{\frak h}^{\nu\mu}$ being the  {\it superpotential},
which is constructed from the densities of spin, bispin, vector field $\xi$
and its derivatives \cite{M,MEN,rem1}. For the Einstein-Hilbert
action one obtains up to a factor of 2 a simple expression \begin{equation}
{\frak J}^\mu= -\frac{1}{4\pi}\left(\sqrt{-g}\xi^{[\mu;\nu]} \right)_{,\nu}
\label{K} \end{equation} which was first given by Komar \cite{K}. It is
impossible simply to ``renormalize'' ${\frak J}^\mu$ by factor of 2 because
the normalization for energy-momentum and angular momentum differ by a
factor of 2. The resolution of this problem is known \cite{Wald}. One needs
to add to the Einstein-Hilbert action $I$ a surface term $I_s$ and apply
Noether's theorem to $I + I_s$. Komar's expression provides a fully
satisfactory notion of the total mass in stationary, asymtotically flat
spacetimes.

For an asymptotically flat spacetime the group of asymptotic
symmetries is an infinite-parametric one. It contains an unique
four-parametric translation subgroup and infinite-parametric subgroup of
supertranslations.  Both groups are the normal subgroups and the Lorentz
group occurs as a factor group of the asymptotic symmetries group by the
infinite dimensional subgroup of supertranslations. {\em Therefore there
does not exist a canonical way of choosing the Poincar\'e group as a
subgroup of the group of asymptotic symmetries. There are too many
Poincar\'e subgroups, one for each supertranslations which is not
translation.} These circumstances generate the main difficulties in the
numerous attempts to find the correct definition of angular momentum.

In our article a new quasigroup approach to the conservation laws
developed in \cite{MEN,N,N1,N2,N3} is applied to asymptotically flat
spacetime. The Poincar\'e quasigroup at future null infinity ($\cal J^+$)
is introduced and compared with other definitions of asymptotic symmetries
that have appeared in the literature.  We define the Noether charge
associated with any element of the Poincar\'e quasialgebra. It may be
regarded as a form of linkages by Tamburino and Winicour \cite{TW1} but
with the new gauge conditions for asymptotic symmetries.

{\em Quasigroups of transformations.} The definition of the quasigroup of
transformations first was given by Batalin \cite{Bat}. Below we outline the
main facts from the theory of the smooth quasigroups of transformations
\cite{rem2}.

Let $\frak M$ be a n-dimensional manifold and the continuos law
of transformation is given by $x'=T_a x , \quad  x\in {\frak M}$, where
\{$a^i$\} is the set of real parameters, $i=1,2,\dots,r$. The set of
transformations $\{T_a\}$ forms a $r$-parametric quasigroup of
transformations (with right action on ${\frak M}$), if:\\
1) there exists a
unit element which is common for all $x^\alpha$ and corresponds to
$a^i=0:\;T_a x|_{a=0}=x$;\\
2) the modified composition law holds:
\[
T_aT_b x
=T_{\varphi(b,a;x)} x;
\]
3) the left and right units coincide:
\[
\varphi(a,0;x)=a, \quad \varphi(0,b;x)=b;
\]
4) the modified law of associativity is satisfied:
\[
 \varphi(\varphi(a,b;x),c;x)=\varphi(a,\varphi(b,c;T_a x);x);
\]
4) the transformation inverse to $T_a$ exists: $x=T^{-1}_a x'$.

The generators of infinitesimal transformations
\[
\Gamma_i=(\partial(T_ax)^\alpha/\partial a^i)|_{a=0}
\partial/\partial x^\alpha \equiv R^\alpha_i\partial/\partial x^\alpha
\]
form {\it quasialgebra} and obey the commutation relations
\begin{equation}
[\Gamma_i, \Gamma_j] =C^p_{ij}(x)\Gamma_p,
\label{eq1}
\end{equation}
where $C^p_{ij}(x)$ are the structure functions satisfying the modified
Jacobi identity
\begin{eqnarray}
C^p_{ij,\alpha}R^\alpha_k + C^p_{jk,\alpha}R^\alpha_i
+C^p_{ki,\alpha}R^\alpha_j   \nonumber \\
+ C^l_{ij}C^p_{kl} + C^l_{jk}C^p_{il} + C^l_{ki}C^p_{jl}=0 .
\label{eq2}
\end{eqnarray}

{\bf Theorem.} Let the given functions $R^\alpha_i, \;C^p_{kj}$
obey the equations (\ref{eq1}), (\ref{eq2}), then locally the quasigroup
of transformations is reconstructed as the solution of set of differential
equations:
\begin{eqnarray}
\frac{\partial\tilde x^\alpha}{\partial a^i}
= R^\alpha_j(\tilde x)\lambda^j_i(a;x),\quad
\tilde x^\alpha(0)=x^\alpha, \label{eq3} \\
\frac{\partial\lambda^i_j}{\partial a^p}
-\frac{\partial\lambda^i_p}{\partial a^j}
+ C^i_{mn}(\tilde x)\lambda^m_p\lambda^n_j=0,
\quad \lambda^i_j(0;x)=\delta^i_j \label{eq4}.
\end{eqnarray}
Eq. (\ref{eq3}) is an analog of the Lie equation, and Eq. (\ref{eq4}) is
the generalized Maurer-Cartan equation.

The notion of quasigroups is not widely known, therefore we give here two
elementary examples playing the important role in the description of
generalized coherent states for the groups $SU(2)$, $SU(1,1)$ and Thomas
precession \cite{NS1,NS2}.

{\em Quasigroup QS(2).} This quasigroup is associated with the group
$SU(2)$ and its action on the two-sphere $S^2$. The sphere $S^2$ admits a
natural quasigroup structure, namely, $S^2$ is a local two-parametric loop
$QS(2)$ which is defined as follows. Let ${\Bbb C}$ be a complex plane and
the isomorphism between points of the sphere and the complex plane ${\Bbb
C}$ is established by the stereographic projection from the north pole of
the sphere:  $\zeta=e^{\rm i\varphi}\tan(\theta/2)$. The nonassociative
multiplication $\star$ is defined by
\[
\zeta\star\eta=\frac{\zeta+\eta}{1-\overline\zeta\eta},
\quad \zeta,\eta\in{\Bbb C}.
\]
where a bar denotes complex conjugation.

{\em Loop QH(2).} This loop is associated with the group
$SU(1,1)$ and its action on the two-dimensional unit hyperboloid $H^2$.
Let $D\subset\Bbb C$ be the open unit disk, $D=\{\zeta\in\Bbb
C:\left|\zeta\right|<1 \}$. We define the binary
operation $\ast$ as following:
\[
\zeta\ast\eta=\frac{\zeta+\eta}{1+\bar\zeta\eta}, \quad\zeta,\eta\in D.
\]
Inside $D$ the set of complex numbers with the operation $\ast$ forms
two-sided loop $QH(2)$, which is isomorphic to the geodesic loop of
two-dimensional Lobachevskii space realized as the upper part of
two-sheeted unit hyperboloid. The isomorphism is established by
$\zeta=e^{\rm i\varphi}\tanh(\theta/2)$ , where $(\theta,\varphi)$ are
inner coordinates on $H^2$.

{\em Null infinity and Poincar\'e quasigroup.} It is known that for
asymptotically flat at future null infinity ($\cal J^+$) spacetime the
group of asymptotic symmetries is the infinite-parametric Newman-Unti (NU)
group which contains the infinitedimensional Bondi-Metzner-Sachs (BMS)
group preserving strong conformal geometry of $\cal J^+$
\cite{NU,BMS,S,S01}.

NU group is the group of tranformations $\cal J^+ \rightarrow \cal J^+$:
\begin{eqnarray}
u \rightarrow u' = f(u,\zeta,\bar \zeta) ,\quad
{\partial f}/{\partial u}> 0, \nonumber \\ \zeta \rightarrow \zeta' =
(\alpha\zeta+\beta)/(\gamma\zeta+\delta).
\quad \alpha\delta -\beta\gamma = 1, \nonumber
\end{eqnarray}
where $f(u,\zeta,\bar \zeta)$ is an arbitrary function, $\zeta$ being a
complex stereographic coordinates at two-dimensional space-like cross
sections of $\cal J^+$ labeled by a coordinate $u$ and its metric is given
by
\[
ds^2=\frac{2d\zeta d\bar\zeta}{|P(u,\zeta,\bar\zeta)|^2},
\]
where we assume $P=VP_0$, $P_0=(1/\sqrt 2)(1+\zeta\bar\zeta)$.
When $V=1$ this metric is reduced to the Bondi metric.

The infinite-parameter BMS-subgroup of NU-group is determined as following
\begin{eqnarray}
u \rightarrow u' = K(\zeta,\bar \zeta)(u+a(\zeta,\bar \zeta)),
 \nonumber \\
\zeta \rightarrow \zeta' = (\alpha\zeta+\beta)/(\gamma\zeta+\delta),
\quad \alpha\delta - \beta\gamma = 1, \nonumber
\end{eqnarray}
where $K(\zeta,\bar \zeta)=(1+\zeta\bar \zeta)/(|\alpha\zeta+\beta|^2 +
|\gamma\zeta+\delta|^2)$ and $a(\zeta,\bar \zeta)$ is an arbitrary
regular function on $S^2$. The infinite-parameter normal subgroup of
BMS-group
\[
\zeta' = \zeta ,\quad u' = u+a(\zeta,\bar \zeta),
\]
is called the subgroup of {\em supertranslations} and contains a
four-parameter normal translation subgroup
\[
a=\frac{A+B\zeta + \bar B \bar\zeta + C \zeta\bar\zeta}{1+\zeta\bar\zeta}.
\]
The factor group of the BMS-group by the supertranslations consists from
the conformal transformations  $S^2 \rightarrow S^2$ and it is isomorphic
to the {\it proper orthochronous Lorentz group}.

On $\cal J^+$ a general element of NU-algebra is given
by (We use the spin coefficients formalism by Newman and Penrose
\cite{PR1} choosing $\kappa=\varepsilon+\bar\varepsilon=0,\; \tau=\bar\pi=
\bar\alpha+\beta,\; \rho=\bar\rho$)
\begin{equation}
\xi=B(u,\zeta,\bar\zeta)\Delta^0 +
C(u,\zeta,\bar\zeta)\bar\delta^0 + \bar C(u,\zeta,\bar\zeta) \delta^0,
\label{eqG}
\end{equation}
where $\eth C=0$ and $\eth,\Delta^0 $, $\delta^0 $ are the standard NP
operators ``edth'', $\Delta$ and  $\delta$ restricted on $\cal J^+$.

In an arbitrary coordinate system on $\cal J^+$ the generators of
four-parameter translation subgroup is given by
\begin{equation}
\xi_a=B_a(\zeta,\bar\zeta)\Delta^0, \quad (C=0),
\end{equation}
where the function $B_a$ is the solution of the following equation
\[
\eth^2 B_a=B_a(4({\bar\alpha}^0)^2 - 2 \eth\bar\alpha^0),
\]
($a$ runs from 1 to 4).
The generators of ``Lorentz group''is determined as following:
\begin{eqnarray}
\xi_A= B_A(u,\zeta,\bar\zeta)\Delta^0
+C_A\bar\delta^0+ \bar C_A \delta^0, \quad (\eth C_A=0),
\label{Q}
\end{eqnarray}
where the function $B_A$ satisfies ($A$ runs from 1 to 6)
\[
\dot B_A = Q_A(u,\zeta,\bar\zeta) +(1/2)(\bar\eth C_A + \eth \bar
C_A),
\]
here $Q_A(u,\zeta,\bar\zeta)$ is an arbitrary real function and  a dot
being derivative with respect to the retarded time $u$.
The generators of the NU-group obey the commutation relations:
\begin{eqnarray}
[ \xi_a ,\xi_b ]=0, \quad
[\xi_a,\xi_B]=C^b_{aB}(u,\zeta,\bar\zeta)\xi_b, \nonumber \\
\left [ \xi_A ,\xi_B \right]
= C^D_{AB}(u,\zeta,\bar\zeta)\xi_D,
\label{str}
\end{eqnarray}
where
$C^b_{aB},\;C^D_{AB}$ are the {\it structure functions}, depending on
{\em arbitrary} function $Q$ \cite{Com1}.

{\em Reduction of NU-group to the Poincar\'e quasigroup}. The point is to
obtain conditions on an arbitrary function $B_A$ in the definition
(\ref{Q}) of the generators of ``Lorentz group''  using the asymptotic
conditions near the ${\cal J}^{+}$. The scheme consists of two steps
 \cite{N1,N3}:

(i) Propagate the asymptotic generators $\xi$ inward along the null
superface $\Gamma$ intersecting $\cal J^+$ in $\Sigma^{+}$ by means of
geodesic deviation equation
\begin{equation}
\nabla^2_l\xi + R(\xi,l)l=0,
\label{Jac}
\end{equation}
imposing the following conditions at $\cal J^+$:
\begin{eqnarray}
\lim_{r\rightarrow\infty}l^\mu l^\nu\pounds_{\xi}g_{\mu\nu}= 0,\quad
\lim_{r\rightarrow\infty}rl^\mu m^\nu\pounds_{\xi}g_{\mu\nu}= 0,\nonumber \\
\lim_{r\rightarrow\infty}l^\mu n^\nu\pounds_{\xi}g_{\mu\nu}
= Q(u,\zeta,\bar\zeta), \nonumber
\end{eqnarray}
$\pounds_\xi$ being Lie derivative with respect to $\xi$.

(ii) Use the commutation relations
\begin{eqnarray}
[ \xi_a ,\xi_b ]=0, \quad
[\xi_a,\xi_B]=C^b_{aB}(r,u,\zeta,\bar\zeta)\xi_b, \nonumber \\
\left [ \xi_A ,\xi_B \right]
= C^D_{AB}(r,u,\zeta,\bar\zeta)\xi_D,
\end{eqnarray}
and asymptotic expansion of $C^a_{bc}$
\[
C= C_0 + C_{1}r^{-1} + C_{2}r^{-2} + \cdots ,
\]
$r$ being a canonical parameter and $C_0$-s the same as in Eq.(\ref{str}),
evaluate all coefficients of these series. The computation yields
\cite{N1}:

(a) The generators of translations are given by
\begin{equation}
\xi_a =B_a\Delta^0,
\label{eqTr}
\end{equation}
where $B_a=l_a(\zeta,\bar\zeta)/|V(u,\zeta,\bar\zeta)|$, $l_a$
satisfies $\eth_0^2 l_a=0$ and a ``standard edth'' $\eth_0$ is referred to
Bondi frame. There are four independent solutions of this equation.

(b) The generators of boosts and rotations are given by
\begin{equation}
\xi_A =B_A\Delta^0 + C_A\bar\delta^0+ \bar C_A \delta^0,
\label{eqBR}
\end{equation}
where $B_A =\Re \tilde B_A$, and the complex function $\tilde B_A$ satisfies
\begin{equation}
\eth^2 \tilde B_A - \tilde B_A \bar\lambda^0 =\frac{\sigma^0}{2}
(3 \eth\bar C_ A - \bar \eth C_A)
+ \bar C_A \eth\sigma^0 + C_A \bar\eth\sigma^0,
\label{eta}
\end{equation}
$\sigma^0$ being the asymptotic shear, $\lambda^0$  the news function,
and $\eth C_A=0$. There are six independent solutions of this equation
which can be written as $C_A = l_A(\zeta,\bar\zeta)/\bar
V(u,\zeta,\bar\zeta)$, where $l_A$ is a solution of the equation $\eth_0
l_A=0$. Further it is convenient to present the function $B_A$ as
\begin{equation}
B_A=\eth\eta\bar C_A + \frac{u-\eta}{2}\eth\bar C_A  + c.c.,
\end{equation}

(c) The generators of the Poincar\'e quasigroup obey at ${\cal J}^{+}$ the
commutation relations:
\begin{eqnarray}
[ \xi_a ,\xi_b ]=0, \quad
[\xi_a,\xi_B]=C^b_{aB}(u,\zeta,\bar\zeta)\xi_b, \nonumber \\
\left [ \xi_A ,\xi_B \right]
= C^D_{AB}(u,\zeta,\bar\zeta)\xi_D, \nonumber
\end{eqnarray}
$C^D_{AB},\;C^i_{aB}$  being the {\it structure functions}, depending on
$B_A = \Re \tilde B_A$, where  $\tilde B_A$ is the solution of the equation
(\ref{eta}). The last commutation relations mean that the generators
of Lorentz quasigroup form a closed algebra.

Note that the same results (a) - (b) can be obtained using instead
of (ii) the asymptotic Killing equations $\pounds_\xi g_{\mu\nu}=
0(1/r^n)$, where the positive integer $n$ depends on the choice of
components. This is the usual way emerging the NU (or BMS) group (see,
{\em e.g.}, \cite{NT} and refrences therein). More exactly these equations
read
\begin{eqnarray}
n^\mu n^\nu \pounds_{\xi}g_{\mu\nu}= O(r), \quad
m^\mu n^\nu \pounds_{\xi}g_{\mu\nu}= O(r), \nonumber \\
m^\mu m^\nu \pounds_{\xi}g_{\mu\nu}= O(r^0), \quad
m^\mu \bar m^\nu \pounds_{\xi}g_{\mu\nu}= O(r^{-2}),
\label{Kil}
\end{eqnarray}
If involve the next orders in powers of $r$, the system (\ref{Kil}) does
not have solution, without additional assumptions. It can be
understood from the following observation:  existence of the solution of
the Killing equtions for an arbitrary order in powers of $r$ implies that
the spacetime has the isometries. Solutions of the geodesic deviation
equation exist in an arbitrary spacetime, while the system of the
asymptotic Killing equations is compatible with (i), (ii) only for $n\leq
2$ \cite{N1,N3}. Note also, that the Jacobi dields play a fundemantal role
in quasigroup approach to quasilocal conservation quantities
\cite{MEN,N,N1,N3}.

As known a group of isometries can be defined as a group which transforms
an arbitrary geodesic to a geodesic one and the Killing vectors satisfy the
geodesic deviation equation for any geodesic.  In the construction above
only {\em null} geodesics passing inward are transformed to the geodesics
under the transformations of the Poincar\'e quasigroup. Besides, the
Poincar\'e {\em quasigroup} is appeared only in a {\em radiative} case.
For non-radiating at $\cal J^+$ systems the generators (\ref{eqTr}),
(\ref{eqBR}) coincide with these ones obtained by Moreschi \cite{Mor}
via solution of the twistor equation on $\cal J^+$, and the Poincar\'e
quasigroup is isomorphic to the Poincar\'e group. It conforms to the
well known results on the reduction of the BMS group to the Poincar\'e
group for the asymptotically flat stationary spacetime.

{\em Isometries compatible with gravitational radiation and the
Poincar\'e quasigroup.} There are known spacetimes with double
(boost-axial) symmetries which are asymptotically flat and admit
radiation \cite{BSw,BS}. Let us introduce on $\cal J^+$ the NP basis
\begin{eqnarray}
n|_{\cal J^+}\equiv \Delta^0=\partial/\partial u, \nonumber \\
m|_{\cal J^+}\equiv \delta^0=(1/\sqrt 2)\left(\partial/\partial\theta
-(i/\sin\theta)\partial/\partial\varphi \right).  \nonumber
\end{eqnarray}
It is easy to show that for the rotational Killing vector (\ref{eta}) is
identity. For the boost Killing vector along the axis $z$ Eq.(\ref{eta})
leads to the following constraint on the news function \cite{N1,N3}:
\begin{equation}
u\cot\theta\frac{\partial \bar\lambda^0}{\partial u}  +
\frac{\partial\bar\lambda^0}{\partial\theta}
+ 2{\bar\lambda}^0 \cot\theta =0.
\end{equation}
This equation is exactly the differential equation for the news function
(Eq.(58) in \cite{BS}), and hence the quasigroup approach conforms to the
known data on the radiative structure of boost-axial spacetime.

{\em Energy-momentum and angular momentum on $\cal J^+$}. As known the
Komar integral is not invariant under a change of the choice of the
generators of time translations in the equivalence class associated with
the given BMS translation. Besides, the resulting energy would not be the
monotonically decreasing Bondi energy but the less physical Newman-Unti
energy \cite{NU}. For an asymptotically flat at future null infinity
spacetime the modified ``gauge invariant'' Komar integral (linkage)
\begin{eqnarray}
L_\xi(\Sigma)=-\lim_{\Sigma_\alpha\rightarrow{\cal J^{+}}}\frac{1}{4\pi}
\oint_{\Sigma_\alpha} \left( \xi^{\left [\alpha;\beta \right]} +
\xi^\rho_{;\rho} l^{\left [\alpha \right.} n^{\left .\beta \right ]}\right)
ds_{\alpha\beta},
\end{eqnarray}
where $\{\Sigma_\alpha\}$ being one-parameter family of spheres, was
introduced by Tamburino and Winicour \cite{TW1}. We adopt this as our
definition of the conserved quantities on $\cal J^+$ associated with the
generators of the Poincar\'e quasigroup. The computation leads to the
following coordinate independent expression \cite{N1,N3}:
\begin{eqnarray}
L_\xi=-(1/4\pi)\Re
\oint\left[B\left(\Psi^0_2 + \sigma^0\lambda^0 -\eth^2\bar\sigma^0\right)
\right.\nonumber \\ \left.+ 2 \bar C\left(\Psi^0_1 +
\sigma^0\eth\bar\sigma^0 +(1/2)\eth(\sigma^0\bar\sigma^0)\right)\right]
d\Omega,\; (\eth C =0).
\end{eqnarray}

The integral four-momentum is given by
\begin{equation}
P_i=  -(1/4\pi)\Re \oint\left[B_i\left(\Psi^0_2 + \sigma^0\lambda^0
-\eth^2\bar\sigma^0\right) \right] d\Omega,
\end{equation}
where $B_i=l_i/{|V|}$ and four-vector
\[
l=\frac{1}{1+|\zeta|^2}\left(1+|\zeta|^2, \zeta + \bar\zeta,
i(\zeta - \bar\zeta), |\zeta|^2 - 1 \right).
\]
Using the Bianchi identities we compute the loss of energy-momentum
\begin{equation}
\dot P_i = -(1/4\pi) \oint B_i |{\cal N}|^2 d\Omega,
\label{P}
\end{equation}
where ${\cal N} = \lambda^0 + 2 \bar\eth\alpha^0 - 4(\alpha^0)^2$. In the
special coordinates ($V=\bar V$) the expression (\ref{P}) coincide with the
expression obtained in \cite{LMN} and in the Bondi frame we
get the standard expression (see, {\it e. g.} \cite{W})
\begin{equation}
\dot P_i = -(1/4\pi) \oint l_i |\lambda^0|^2 d\Omega.
\end{equation}

The angular momentum is given by
\begin{equation}
M_A=-(1/4\pi)\Re \oint \bar l_A {\cal F}d\Omega,
\label{AM}
\end{equation}
where $l_A$ is the solution of $\eth l_A = 0$ and
\begin{eqnarray}
{\cal F}=2\Psi^0_1 + 2\sigma^0\eth\bar\sigma^0 + \eth(\sigma^0\bar\sigma^0)
+3\eth\eta(\Psi^0_2 + \sigma^0\lambda^0 - \eth^2\bar\sigma^0)
\nonumber \\
+ (\eta - u)\left(\eth\Psi^0_2 + \eth(\sigma^0\lambda^0)
-\eth^3\bar\sigma^0\right). \nonumber
\end{eqnarray}
It yields
\begin{eqnarray}
\dot M_A = -(1/4\pi)\Re \oint \bar l_A
\frac{\partial}{\partial u}\left(\frac{{\cal F}}{V |V|^2} \right)
V |V|^2 d\Omega .
\end{eqnarray}

Let $\Sigma_1$ and $\Sigma_2$ be arbitrary cross-sections at $\cal J^+$, then
the flux of the energy-momentum and angular momentum is given by
\begin{eqnarray}
P_i(\Sigma_2) - P_i(\Sigma_1) = -(1/4\pi) \oint B_i |{\cal N}|^2 d\Omega du,\\
M_A(\Sigma_2) - M_A(\Sigma_1) \nonumber \\
= -(1/4\pi)\Re \oint \bar l_A
\frac{\partial}{\partial u}\left(\frac{{\cal F}}{V |V|^2} \right)
V |V|^2 d\Omega du,
\end{eqnarray}
where the integration is performed over the domain $\Omega \subset
{\cal J^+}$ contained between $\Sigma_1$ and $\Sigma_2$.

Geroch and Winicour \cite{GW} have given a list of properties which
conserved quantities  $P(\xi,\Sigma)$ defined at $\cal J^+$ should have:\\
(1) $P(\xi,\Sigma)$ should be linear in the generators of the asymptotic
symmetry group.\\
(2) $P(\xi,\Sigma)$ should be invariant with respect to the conformal transformations
$\tilde g_{\mu\nu}=\tilde\Omega^2 g_{\mu\nu}$. \\
(3) The expression $P(\xi,\Sigma)$ should depends on the geometry of
$\cal J^+$  and behavior of generators in the neighbourhood of $\cal J^+$.\\
(4)  $P(\xi,\Sigma)$ should be proportional to the corresponding Komar integral for the
exact symmetries and coincide with the Bondi four-momentum when $\xi$ is a
BMS translation. \\
(5) $P(\xi,\Sigma)$ should be define also for the system with radiation on $\cal
J^+$.\\
(6) There should exist a flux integral $\cal I$ which is linear in
$\xi$ and which gives the difference $P(\xi,\Sigma') - P(\xi,\Sigma)$, for
$\Sigma'$ and $\Sigma'$ closed two-surfaces on $\cal J^+$. \\
(7) In Minkowski spacetime $P(\xi,\Sigma)$ should vanish identically.

Our definition of the conserved quantities is free from the supertranlation
ambiguity and satisfies all conditions (1) - (7) \cite{N1}. In the
linearized theory, for the non-contorted two-surfaces, the expression
(\ref{AM}) coincide with this one obtained by Bramson \cite{Br}.

\acknowledgements
I would like to thank N. V. Mitskievich, L. V. Sabinin, T. Zannias, R. D.
Sorkin and D. E. Sudarsky for helpfull discussions and comments. This work was supported by CONACyT, Grant No. 1626P-E.



\begin{references}
\bibitem{rem2} In fact Batalin introduced not a quasigroup but a loop of
transformations. The theory of smooth loops generalizing the Lie theory was
developed by Sabinin. \cite{S1,S2,S3}
\bibitem{Com1} Note, that the last commutation relation can be written
as $\left [ \xi_A ,\xi_B \right]=C^D_{AB}\xi_D
+ C^a_{AB}(u,\zeta,\bar\zeta)\xi_a,$ where $C^D_{AB}$ are the {\it
structure constants} of the Lorentz group and
$C^a_{AB}(u,\zeta,\bar\zeta)$ the {\it structure functions} depending
on $Q$.
\end{references}
\end{document}